# Pushing an Altermagnet to the Ultimate 2D Limit: Symmetry Breaking in Monolayers of GdAlSi

*Oleg E. Parfenov, Dmitry V. Averyanov, Ivan S. Sokolov, Alexey N. Mihalyuk, Oleg A. Kondratev, Alexander N. Taldenkov, Andrey M. Tokmachev, and Vyacheslav G. Storchak\**

O. E. Parfenov, D. V. Averyanov, I. S. Sokolov, O. A. Kondratev, A. N. Taldenkov, A. M. Tokmachev, V. G. Storchak
National Research Center "Kurchatov Institute", Kurchatov Sq. 1, 123182 Moscow, Russia
E-mail: vgstorchak9@gmail.com

A. N. Mihalyuk
Institute of High Technologies and Advanced Materials, Far Eastern Federal University, 690950 Vladivostok, Russia
Institute of Automation and Control Processes FEB RAS, 690041 Vladivostok, Russia

Keywords: altermagnetism, 2D material, anomalous Hall effect, Weyl semimetal, GdAlSi

Altermagnets have emerged as a class of materials combining certain ferromagnetic properties with vanishing net magnetization. This combination is highly promising for spintronics, especially if a material can be brought to a nanoscale size. However, experimental studies of the 2D limit of altermagnets and evolution of their properties with thickness are lacking. Here, we study epitaxial films on silicon of the Weyl altermagnet GdAlSi ranging from more than a hundred unit cells to a single unit cell. The films are synthesized by molecular beam epitaxy and, expectedly, do not show any discernible net magnetic moments. Electron transport studies reveal a remarkable transformation of the electron state with the film thickness. Thick films exhibit negative longitudinal magnetoresistance associated with the chiral anomaly but do not demonstrate altermagnetic properties in electron transport due to symmetry restrictions. In ultrathin films, a spontaneous anomalous Hall effect manifests itself, indicating a non-relativistic spin splitting in the electronic structure. The transformation is associated with crystal symmetry breaking accompanying the 3D-to-2D crossover. The work highlights the role of dimensionality in altermagnetism and provides a platform for studies of altermagnets aiming at ultra-compact spintronics.

# 1. Introduction

A new elementary class of magnetic materials, altermagnets (AM), has recently emerged and is currently at the forefront of research on magnetism.[1-4] These materials combine vanishing net magnetization, as antiferromagnets (AFM), with spin-polarization phenomena related to time reversal symmetry breaking, typical of ferromagnets (FM). The apparent merge of FM and AFM characteristics stems from compensated collinear moments connected by crystal-rotation symmetries: alternating local structures around spins produce spin splitting in reciprocal space,[1,5,6] evidenced by angle-resolved photoemission spectroscopy (ARPES) with spin resolution.[7,8] The non-relativistic origin of the splitting is conducive to strong spin-based responses. The combination of FM features with AFM-like vanishing stray fields and THz dynamics constitutes a key advantage of the new class of materials. AMs are attractive for spintronics, ultrafast optics, magnonics, studies of neuromorphic systems. The anomalous Hall effect (AHE) is a hallmark of altermagnetism, demonstrated in Refs. [9-12] and associated with the anisotropic magnetic dipole.[13] AMs exhibit the anomalous Nernst effect,[14,15] efficient interconversion of charge and spin,[16,17] THz emission induced by the inverse spin splitting effect.[18] Significant progress has been made in studies of bulk AMs and rather thick, bulk-like films. However, planar nanoelectronics aiming at compact devices requires ultrathin films. The question is how the properties of an AM material evolve as it approaches and reaches the 2D limit.

2D magnetic systems, including those with zero net magnetic moment,[19] are currently exciting significant interest because they promise advances in search for unconventional quantum phases, fundamentals of magnetism, and applications in ultra-compact spintronics.[20,21] 2D magnets are highly amenable to external stimuli such as pressure,[22] gating[23] or magnetic fields;[24] they exhibit strong magnetotransport effects such as colossal lateral[25] and tunneling[26] magnetoresistance (MR). An important feature of 2D magnets is high sensitivity of their properties to the film thickness.[24,27-29] The concept of AM extends to thin films and interfaces.[30] Various candidate AM materials, $RuO_2$,[9,31] $Mn_5Si_3$,[11,12,15] MnSe,[32] MnTe,[7,33] and CrSb,[34] have been produced as films but rather thick, dozens of nm. Only a few experimental studies venture into the region of ultrathin films, down to the thickness of several unit cells.[35,36] In contrast, theoretical research often focuses on monolayer AMs.[37-42] The latter predicts topological structures related to Weyl[39,43] and Majorana[44] fermions, foresees superconductivity,[45] spin-momentum locked transport,[39] various magnetocaloric[46] and piezo[40-42] effects. In studies of AM systems and their properties, symmetry considerations play a crucial role. In particular, crystal symmetry breaking affects functional properties of AM materials and can be employed in their design.[36,41,42,47] This issue is vital for 2D AMs because the crystal structures of systems approaching the 2D limit lose translational symmetry operations and become sensitive to interfacial effects such as tensile strains caused by the substrate. Therefore, it is important to study the evolution of AMs as they transform from bulk to the ultimate limit of ultrathin films.

To enable such a study, certain conditions are wanted. First, a synthetic route for the AM material should be available, to produce epitaxial films of any thickness. The epitaxial quality is important for studies and applications of crystal-structure-related effects.[2] Second, the 3D-to-2D crossover should result in tangible changes to the crystal symmetry and the related properties. Then, the material should be conductive down to the thinnest films, to make a study of electron transport possible. Finally, potential applications would benefit considerably from integration of the AM material with a mature semiconductor platform, such as that of Si. The material landscape of AMs is rather scarce;[4] yet, we can select a material satisfying the prerequisites posed above: our choice is the tetragonal polymorph of GdAlSi (**Figure 1**a), proposed recently as a candidate AM.[48] GdAlSi is a Weyl semimetal, according to both density functional theory (DFT) calculations[48,49] and ARPES measurements.[48] In contrast to other AMs formed by transition metals, the magnetic moments in GdAlSi arise from $f$-shells. Following the notation of Ref. [1] for AMs, the spin-group symmetry of GdAlSi is $[C_2||C_{4z}t]$,[48] as that of $RuO_2$,[1] combining a 2-fold spin-space rotation with a 4-fold real space rotation. The latter symmetry (rotation + translation) is illustrated by transformation of the non-magnetic environment of Gd atoms with different spins – $Si_6$ and $Al_6$ prisms (Figure 1b,c) – essential for the emergence of the non-relativistic spin splitting (NRSS). The NRSS is confirmed by DFT calculations and explained within the framework of multipole analysis.[48] Moreover, GdAlSi is suggested as a functional material for spin transistors and spin valves.[48] However, the GdAlSi properties related to NRSS are yet to be explored experimentally. A recent study on GdAlSi polymorphism demonstrates that epitaxial films of the tetragonal GdAlSi can be synthesized directly on silicon.[50] This development opens an opportunity for thickness-dependent studies of the material properties down to a single unit cell.

Here, we produce and study epitaxial films of GdAlSi on silicon, ranging from bulk-like ($d$ = 193 nm) to 1 unit cell thick films, down to the ultimate 2D limit of the compound. We demonstrate the absence of any detectable net magnetic moment irrespective of the film thickness. A dramatic transformation is found in the electron transport properties related to the crystal symmetry breaking caused by the 3D-to-2D crossover in the material dimensionality. In particular, we track the disappearance of the chiral anomaly, a hallmark of Weyl physics, and the emergence of spontaneous AHE attributed to NRSS.

## 2. Results and Discussion

### 2.1. Synthesis and Characterization of GdAlSi Films

Synthesis of epitaxial GdAlSi films of any arbitrary thickness is challenging. First, the Gd-Al-Si system exhibits a large variety of phases with different stoichiometries, from binary $GdSi_x$ and $GdAl_x$ to ternary $GdAl_2Si_2$, $Gd_4AlSi_3$ or $Gd_6Al_3Si$. Therefore, the synthesis conditions and the amounts of reactants should be carefully selected to avoid unwanted side products. Second, tetragonal GdAlSi is not the most stable polymorph of GdAlSi in ultrathin films – below some critical thickness (around 8 nm), the trigonal polymorph of GdAlSi becomes the ground state.[50] Therefore, the material requires stabilization by the substrate. To this end, we employ the (001) surface of silicon matching the symmetry of the tetragonal polymorph of GdAlSi. This approach is inspired by studies of $EuSi_2$, a compound isoelectronic to GdAlSi: its tetragonal polymorph, isomorphous to that of GdAlSi, is stabilized on Si(001)[51] whereas a trigonal polymorph is grown on Si(111).[28,52] Synthesis of GdAlSi on Si(001) carries other benefits: (i) the substrate becomes one of the reactants, simplifying realization of the precise stoichiometry because then only the Gd and Al fluxes require matching; (ii) the resulting material is seamlessly integrated with silicon. Besides the general synthetic route, the technique of GdAlSi synthesis is also of importance because it affects strongly the quality of the films. Our choice here is molecular beam epitaxy (MBE), a technique that has proven useful in studies of ultrathin films of Gd-based materials on silicon.[25,28,29,50]

The synthesis starts with removal of the surface oxide from the substrate surface. This process is carried out thermally in the vacuum conditions of the MBE chamber. The bare Si(001) surface exposes its characteristic $(1 \times 2) + (2 \times 1)$ reconstruction. To synthesize GdAlSi, Gd and Al are co-deposited on the bare substrate. The reaction proceeds at relatively mild conditions. The substrate temperature is kept at 400 °C, well below $T = 1000$ °C used for synthesis of bulk GdAlSi.[48,49,53] Another difference is that large excess of Al is utilized in synthesis of bulk GdAlSi[48,49,53] whereas synthesis of the films employs an excess of available Si supplied from the substrate *via* the vacancy mechanism. The thickness of the films is controlled by the deposition time. This way, films of different thickness, down to a single monolayer (ML; for convenience, we call a monolayer the film thickness of one unit cell), are synthesized. GdAlSi may be oxidized by air oxygen. To prevent degradation of the samples, they are capped with a layer of the amorphous nonmagnetic insulator $SiO_x$.

To study the atomic structure of the films, we employed diffraction techniques, reflection high-energy electron diffraction (RHEED) and X-ray diffraction (XRD). RHEED probes the surface of the sample in the growth chamber. Figure 1d shows a typical 3D RHEED spectrum of 1 ML GdAlSi. As expected, the spectrum corresponds to the tetragonal polymorph of GdAlSi and signifies the epitaxial growth of the film. The direction *c* of the GdAlSi unit cell is orthogonal to the film surface. RHEED spectra for other GdAlSi films are demonstrated in Figure S1 of the Supporting Information. All the films are epitaxial but the surface becomes rough as the film thickness increases. The capped films are studied *ex situ* by XRD. Figure 1e provides a θ-2θ scan of 5 ML GdAlSi. The scan demonstrates a series of $(0\,0\,4n)$ peaks corresponding to an epitaxial film of GdAlSi without any side phases. The XRD scans for the other films are presented in Figure S2 of the Supporting Information. An important feature of the scans is the presence of thickness fringes around the peaks (see also Figure S3 of the Supporting Information). The 3 ways to determine the film thickness, employing (i) the calibrated time of the metal deposition, (ii) thickness fringes, and (iii) Debye-Scherrer analysis of the XRD peaks, demonstrate a good agreement. The films produced span over a wide range of thickness: from the ultimate 2D limit of 1 ML to bulk-like 134 ML.

**2.2. Bulk Limit of GdAlSi Films**

First, we consider the bulk limit of GdAlSi. We suggest (and show below) that the properties of the thickest film (134 ML) are close to those of the bulk. Magnetic properties of GdAlSi stem from half-filled 4*f* shells of $Gd^{3+}$ ions providing local magnetic moments of 7 $\mu_B$/Gd. The ground-state magnetic configuration of GdAlSi has been established by DFT calculations in Refs. [48,49]. It is shown in Figure 1a. The MBE-grown films have a significant area making possible their studies by SQUID. **Figure 2**a shows temperature dependences of the molar magnetic susceptibility in in-plane magnetic fields. The results are characteristic of the AFM ordering of the magnetic moments and agree with the earlier studies of bulk GdAlSi in Refs. [48,49,53]. In particular, the magnetic transition temperature $T_N$ is about 32 K. Similar dependences in out-of-plane magnetic fields are shown in Figure S4 of the Supporting Information. The magnetic anisotropy is weak which is typical of Gd-based magnets;[54] the same conclusion is made in Ref. [49]. The magnetization in the paramagnetic region fits well by the Curie-Weiss law with the parameter $\theta \approx$ -119 K (Figure S5 of the Supporting Information), in agreement with the data for bulk GdAlSi.[48,49,53]

To get information on the electron states, we studied electron transport in GdAlSi. Figure 2b demonstrates temperature dependence of the film resistance in zero magnetic field and in a magnetic field 9 T of different directions. The film conductivity corresponds to that of a metal. The $R(T)$ dependences display a feature around $T_N$ signifying the decrease of magnetic scattering in the magnetically ordered state. In magnetic fields, this feature shifts toward lower temperature (see Figure S6 of the Supporting Information). Such shift is consistent with an AFM ordering. Figure 2b suggests negative longitudinal MR. The conclusion gains support by studies of MR in different magnetic fields (Figure 2c). This is a signature of non-trivial band structure, commonly ascribed to the chiral anomaly, *i.e.* breakdown of the chiral symmetry in parallel magnetic and electric fields.[55] This observation is not surprising, taking into account that GdAlSi is identified as a Weyl semimetal.[48,49] The chiral anomaly is not directly related to the magnetic state and, therefore, shows up at temperatures above $T_N$ (see Figure S7 of the Supporting Information). In topological semimetals, one expects charge carriers with a low effective mass. The presence of such carriers in GdAlSi is confirmed by analysis of Shubnikov-de Haas oscillations in the Hall resistance (Figure S8 of the Supporting Information). Thus, the 134 ML film of GdAlSi displays salient features of the bulk material, therefore can be considered as the bulk limit.

AMs are expected to exhibit FM effects such as AHE. The Hall effect in 134 ML GdAlSi depends non-linearly on the applied magnetic field. A similar observation in bulk GdAlSi, made in Ref. [49], has prompted a claim of AHE. However, we cannot support this claim for a number of reasons: (i) the non-linearity is not related to the magnetic transition – the effect persists well above $T_N$, up to room temperature; (ii) there is no hysteresis in the magnetic field dependence of $R_{xy}$; (iii) the characteristic magnetic fields are too high. To clarify the situation, we employ a symmetry analysis. The space group of bulk GdAlSi is $I4_1md$ (109); its magnetic space group is $I4_1'm'd$.[48] A spontaneous AHE can occur only if the magnetic point group (MPG) allows a ferromagnetic state.[56,57] In bulk GdAlSi, the MPG, $4'm'm$, is not compatible with ferromagnetism. Therefore, the non-linearity of the Hall effect should be attributed to multiple carrier bands rather than AHE. The same conclusion has been drawn for the candidate AM CrSb which MPG, $6'/m'mm'$, is also not compatible with ferromagnetism.[57] In Ref. [49], the multiband explanation was discarded because of a rather poor fit of the data to a simple two-band model. Indeed, description of the electron transport data requires three types of carriers (for comparison, four types of carriers are required to describe electron transport in CrSb[57]). Figure 2d shows the fit of the longitudinal and transverse magnetoconductance to a three-band model. The model evokes bands of (relatively) heavy holes, light holes and electrons. The carrier mobility of light holes reaches almost 1500 cm$^2$ V$^{-1}$ s$^{-1}$. Temperature dependence of the carrier mobility and concentration shows that the band structure is affected by the magnetic transition (see Figure S9 of the Supporting Information). In summary, AM features of GdAlSi are hidden in the bulk limit; one needs to lower the symmetry to remove the restriction on the AHE observation.

## 2.3. Monolayer Limit of GdAlSi

Reduction of the dimensionality is a natural way to lower the symmetry: transition from the bulk to monolayers results in the loss of translational symmetry along the $c$ axis. Simple symmetry considerations suggest that the MPG of 1 ML GdAlSi is $m'm'2$: in contrast to the bulk, the MPG is compatible with ferromagnetism, lifting the restriction and making the observation of AHE possible. However, the altermagnetism of GdAlSi relies upon the $C_{4z}t$ operation, an operation which is lost in 1 ML GdAlSi. Therefore, it is necessary to perform band structure calculations to identify the characteristics of the bulk that persist in 1 ML and those that are lost. We carried out DFT calculations of bulk and 1 ML GdAlSi. First, we checked the consistency of the results with those reported earlier in Ref. [48] and confirmed that the band structure of bulk GdAlSi is reproduced in our calculations (see Figure S10 of the Supporting Information). **Figures 3**a,b compare the band structures of bulk and 1 ML GdAlSi. Both materials demonstrate NRSS: although 1 ML GdAlSi is, strictly speaking, not an AM, it is still a system with NRSS, as the bulk AM. To get further insights, we consider the band structures of the bulk and 1 ML along two orthogonal directions, $\Gamma - X$ and $\Gamma - X'$ (see Figures 3c,d): the image for bulk GdAlSi is symmetric but that for 1 ML is not, in accordance with the lower crystal symmetry of the 1 ML system. In both cases, we observe the flip of the spin splitting at the $\Gamma$ point;[48] however, in the case of 1 ML, not all the bands follow this pattern.

The loss of the symmetry operation $C_{4z}t$ means that the absolute values of the opposite magnetic moments in GdAlSi are not bound by symmetry to be exactly equal. However, the local magnetic moments of the half-filled 4$f$-shells of Gd$^{3+}$ ions are rather stable, and any significant net magnetization is not expected. We carried out SQUID measurements of 1 ML GdAlSi (see **Figure 4**a). The net magnetic moment in the material is negligible signifying that the local magnetic moments of Gd ions are compensated. The lack of detected magnetic moment in 1 ML GdAlSi is not due to insufficient sensitivity of the SQUID measurements: for illustration, we show that SQUID readily detects the FM state in a film of trigonal GdAlSi of about the same thickness (Figure 4a). The conclusion on the lack of any noticeable magnetic moment in 1 ML GdAlSi is confirmed by $M(H)$ in the two polymorphs (see Figure S11 of the Supporting Information). The change from the bulk to 1 ML is marked by a fundamental transformation in the electron transport. Figure 4b demonstrates a significant AHE in 1 ML GdAlSi, in sharp contrast to the bulk. The crystal symmetry breaking due to the reduction of dimensionality is sufficient to lift the restriction on the AHE observation but the substrate may also add to the symmetry breaking. The AHE gradually disappears with the temperature increase (Figure 4b). A small effect still persists above $T_N$ of GdAlSi (see Figure S12 of the Supporting Information), probably due to magnetic correlations. The AHE demonstrates a hysteretic behavior (Figure 4c), similar to other AMs such as MnTe[9] and $Mn_5Si_3$.[10]

Other transport properties of 1 ML GdAlSi also differ significantly from those in the bulk. For instance, MR in 1 ML depends weakly on the magnetic field direction (Figure 4d), in contrast to a significant anisotropy of MR in the bulk (Figure 2c). The MR in 1 ML is negative. The resistivity follows the logarithmic temperature dependence (Figure 4e). Together with the lack of MR anisotropy, it may point at the Kondo interaction of carriers with local magnetic moments, similar to ultrathin films of $GdSi_2$.[25] The negative MR in 1 ML GdAlSi below 20 K (see Figure 4e and Figure S13 of the Supporting Information) may be associated with the Kondo effect. The AHE in magnetic systems can appear due to different mechanisms.[58] To get insight into the AHE in GdAlSi, we plotted the dependence of the transverse (AHE) conductivity on the longitudinal conductivity (Figure 4f). The plateau in the dependence points at the intrinsic mechanism of the AHE.

## 2.4. Evolution of GdAlSi Films with Thickness

Up to now, we considered two opposite limits, bulk and 1 ML. However, it is instructive to track the changes in the electronic structure and properties between the two limits. **Figure 5**a shows the dependence of the film conductivity on its thickness. The conductivity of 34 ML is about the same as that in 134 ML – both films belong to the bulk limit. At lower thickness, the conductivity decreases sharply, which points at the increasing role of localization effects. Temperature dependence of the resistivity $\rho_{xx}(T)$ in thick films is qualitatively different from that in thin films (see Figure S14 of the Supporting Information). The changes in the electron state with the film thickness are well illustrated by longitudinal MR (Figure 5b). In the bulk, the MR is negative because of the chiral anomaly. In thin films, the chiral anomaly is suppressed. A similar effect has been observed in $SrSi_2$ films[59] and typical Weyl semimetals NbP and TaP.[60] For intermediate thickness, the longitudinal MR is positive but the MR again becomes negative in ultrathin films. The thickness-dependent evolution of the MR in out-of-plane magnetic fields (from positive to negative) is shown in Figures S15-S17 of the Supporting Information.

Figure 5c shows that the Hall effect undergoes significant changes between 10 ML (multiple-band carriers) and 5 ML (AHE) GdAlSi. The non-linear part of the Hall effect even changes its sign. The inverted *S*-shape form at 5 ML is what we expect for AHE in a Gd-based magnet – similar shapes are observed for AHE in the 2D magnets $GdSi_2$,[28,61] $GdGe_2$,[29] Gr/Gd/Si.[62] For ultrathin films, the evolution of the AHE resistance with temperature is shown in Figure S18 of the Supporting Information. The evolution of the MR and AHE can be associated with the changes in the carrier concentration and mobility (Figures 5d,e). In films of thickness below 10 ML, the three-band model becomes irrelevant because we are not able to detect the light carriers. The description by a single type of carriers shows the opposite trends in the carrier mobility and concentration: the former decreases and the latter increases significantly as the film enters the 2D limit. Moreover, the type of the carriers changes between 2 and 1 ML, from heavy holes to heavy electrons. The changes to the carrier state are accompanied by evolution of the MR anisotropy (see Figure 5f and Figure S19 of the Supporting Information). In thick films, the MR anisotropy is significant but below 10 ML, it becomes rather weak. Therefore, we can roughly place the crossover between the electron states into the range of thicknesses from 5 to 10 ML.

## 3. Conclusion

The research on altermagnetism is in its infancy; significant issues are yet to be addressed experimentally. One of the research areas requiring attention is the 2D limit of AMs and thickness-dependent evolution of their properties. Here, we tackled the problem by synthesis and analysis of the candidate Weyl AM GdAlSi. We managed to develop a synthetic route to epitaxial films of GdAlSi directly on silicon. The AHE is a characteristic effect expected in AMs but symmetry restrictions prevent its observation in thick films of GdAlSi emulating the bulk. However, these films exhibit other fascinating properties, such as negative longitudinal MR, ascribed to the chiral anomaly. At a thickness of several ML we detect a crossover to a fundamentally different electron state. The functional properties of this state are what we expect for an AM or, alternatively, an AFM system with NRSS. Crystal symmetry breaking due to a lower dimensionality lifts restriction on the AHE and, indeed, a spontaneous AHE is observed in an ML of GdAlSi. The approach can be extended to other magnetic systems such as various MAB materials related to GdAlSi. The present work paves the way for experimental studies and applications of AMs brought to the 2D limit.

## 4. Methods

*Synthesis*: The molecular beam epitaxy of the GdAlSi films was carried out in a Riber Compact MBE system in the UHV conditions (the base pressure in the MBE chamber did not exceed $10^{-10}$ Torr). To stabilize the tetragonal polymorph, the Si(001) substrate (1 in × 1 in wafers with miscut angles less than 0.5°) was employed. The oxide layer on the Si(001) surface was removed by heating the wafer up to 950 °C; the substrate temperature was established using a PhotriX ML-AAPX infrared pyrometer operating at a wavelength of 0.9 μm. As a result, the characteristic (1 × 2) + (2 × 1) reconstruction of the bare Si(001) surface was formed. The substrate was kept at 400 °C and, then, GdAlSi was synthesized by reaction of Si, delivered from the substrate, with Gd and Al deposited on top of the wafer. 4N Gd and 5N Al were supplied from the corresponding Knudsen cell effusion sources heated up to 1210 and 905 °C, respectively. The pressures of Gd and Al in the MBE synthesis chamber, determined with a Bayard-Alpert ionization gauge, were $1\cdot10^{-8}$ and $6\cdot10^{-9}$ Torr, respectively. GdAlSi films were protected from air oxidation by a 20-nm layer of $SiO_x$, an amorphous inert nonmagnetic insulator deposited at room temperature that does not affect the studies of the films.

*Characterization*: A combination of diffraction techniques was employed to study the atomic structure and quality of the films. The *in situ* studies were carried out using a RHEED diffractometer furnished with the kSA 400 analytical system. RHEED was used to monitor the natural oxide removal and the synthesis of the GdAlSi films. The other measurements were taken *ex situ*, after capping the films with $SiO_x$. θ-2θ XRD scans were produced by a Rigaku SmartLab 9 kW diffractometer using the Cu $K_{\alpha 1}$ radiation (a wavelength of 1.54056 Å). The magnetic properties of the films were studied by an MPMS XL-7 SQUID magnetometer. Samples with a lateral size of 5 mm were placed into plastic straws and oriented with respect to the external magnetic field with an accuracy of better than 2°. The measurements were carried out employing the reciprocating sample option (RSO) mode. The electron transport in the films was studied by a Lake Shore 9709A measurement system. Square samples 5 mm × 5 mm were used to determine the resistivity and Hall effect characteristics of GdAlSi. The electrical contacts to the films were produced by deposition of an Ag-Sn-Ga alloy. The Ohmic behavior of the contacts was established by I-V characteristic curves. Four-contact measurements were carried out according to the ASTM Standard F76.

*Computational Techniques*: The calculations of the electronic structure of GdAlSi were based on the DFT+U computational scheme as implemented in the Vienna *ab initio* simulation package (VASP).[63] The projector augmented wave approach[64] was used to describe the electron-ion interaction. The generalized gradient approximation (GGA) of Perdew, Burke, and Ernzerhof (PBE)[65] was taken as the exchange-correlation functional. The calculations accounted for the spin-orbit coupling (SOC). A Hubbard U of 7.0 eV was applied to the Gd atoms using Dudarev's method[66] to avoid the self-interaction errors arising from an incorrect description of partially filled $f$ states of Gd. A kinetic energy cutoff of 300 eV was used in all the calculations. In the case of 1 ML GdAlSi, the vacuum size along the non-periodic direction was set to 15 Å. The 12×12×4 and 12×12×1 Γ-centered $k$-point meshes were used to sample the 3D and 2D Brillouin zones of the GdAlSi bulk and 1 ML structures, respectively. The geometry of the crystal structure was optimized until the residual forces on the atoms were below 0.001 eV/Å. The resulting lattice parameters $a$ = 4.1311 Å and $c$ = 14.4866 Å are close to the experimental values.[49] The AFM configuration shown in Figure 1a (interlayer AFM ordering with Gd magnetic moments oriented along the $c$-axis) was used for band structure calculations as the energetically most favorable configuration.[48,49]

**Supporting Information**

Supporting Information is available from the Wiley Online Library or from the author.

**Acknowledgements**

This work was supported by NRC "Kurchatov Institute" and the Russian Science Foundation [grants No. 22-13-00004 (synthesis), No. 24-19-00038 (magnetism studies), and No. 20-79-10028 (studies of electron transport)]. D.V.A. acknowledges support from the President's scholarship (SP 3111.2022.5). The measurements were carried out using equipment of the resource centers of electrophysical and laboratory X-ray techniques at NRC "Kurchatov Institute".

Received: ((will be filled in by the editorial staff))
Revised: ((will be filled in by the editorial staff))
Published online: ((will be filled in by the editorial staff))

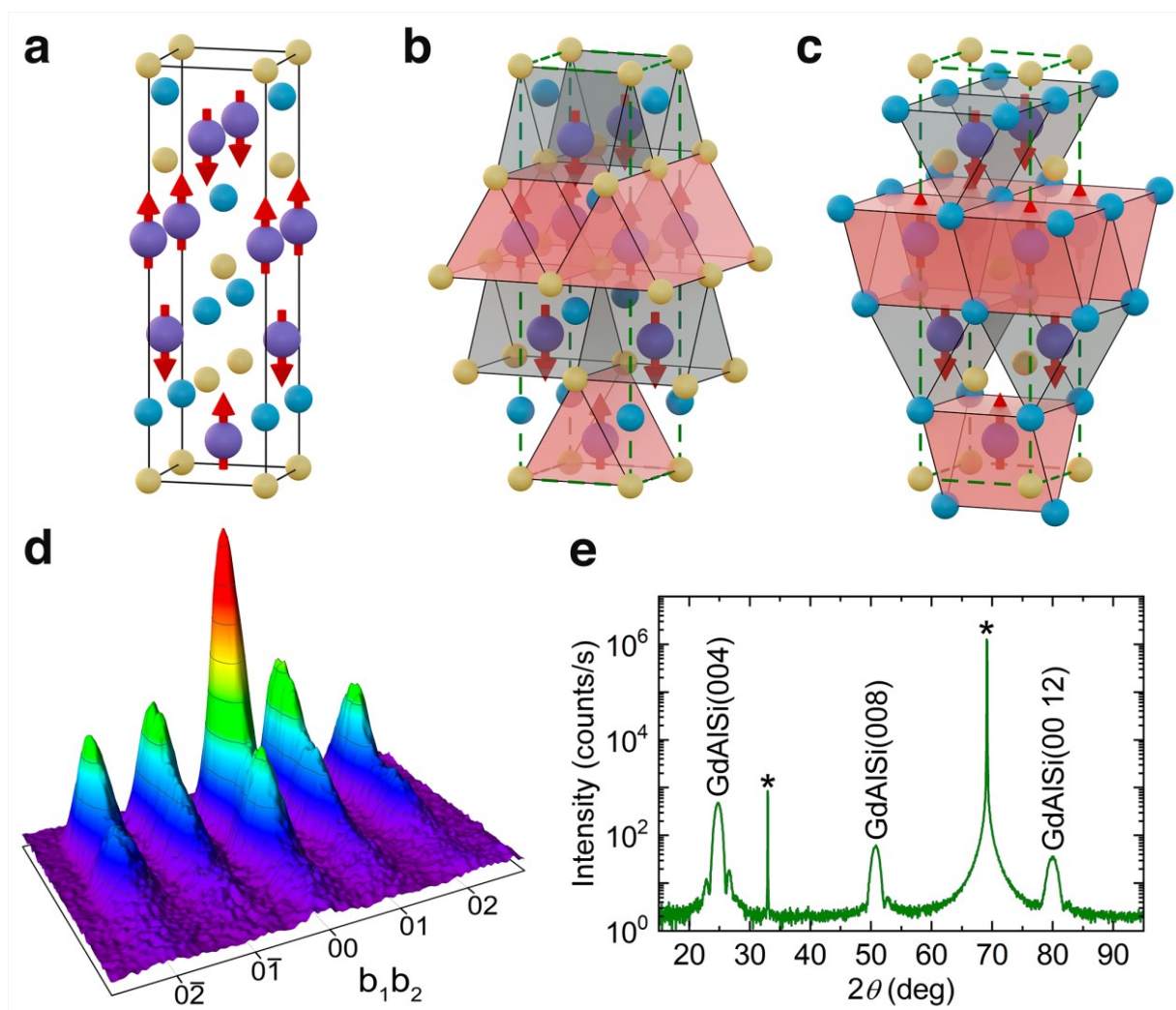

**Figure 1.** Atomic structure of tetragonal GdAlSi. a) Unit cell of GdAlSi comprising 4 formula units; atoms of Gd (purple), Al (blue) and Si (vanilla) are shown; arrows at Gd atoms depict the magnetic configuration of the lowest energy. b) $Si_6$ prisms around Gd atoms demonstrating the $C_{4z}t$ symmetry connecting Gd atoms with opposite spins. c) $Al_6$ prisms around Gd atoms demonstrating the $C_{4z}t$ symmetry connecting Gd atoms with opposite spins. d) 3D RHEED image of 1 ML GdAlSi; the reflexes are marked by Miller-Bravais indices $(b_1b_2)$ for the basal plane. e) θ-2θ XRD scan of 5 ML GdAlSi; asterisks denote peaks from Si(001).

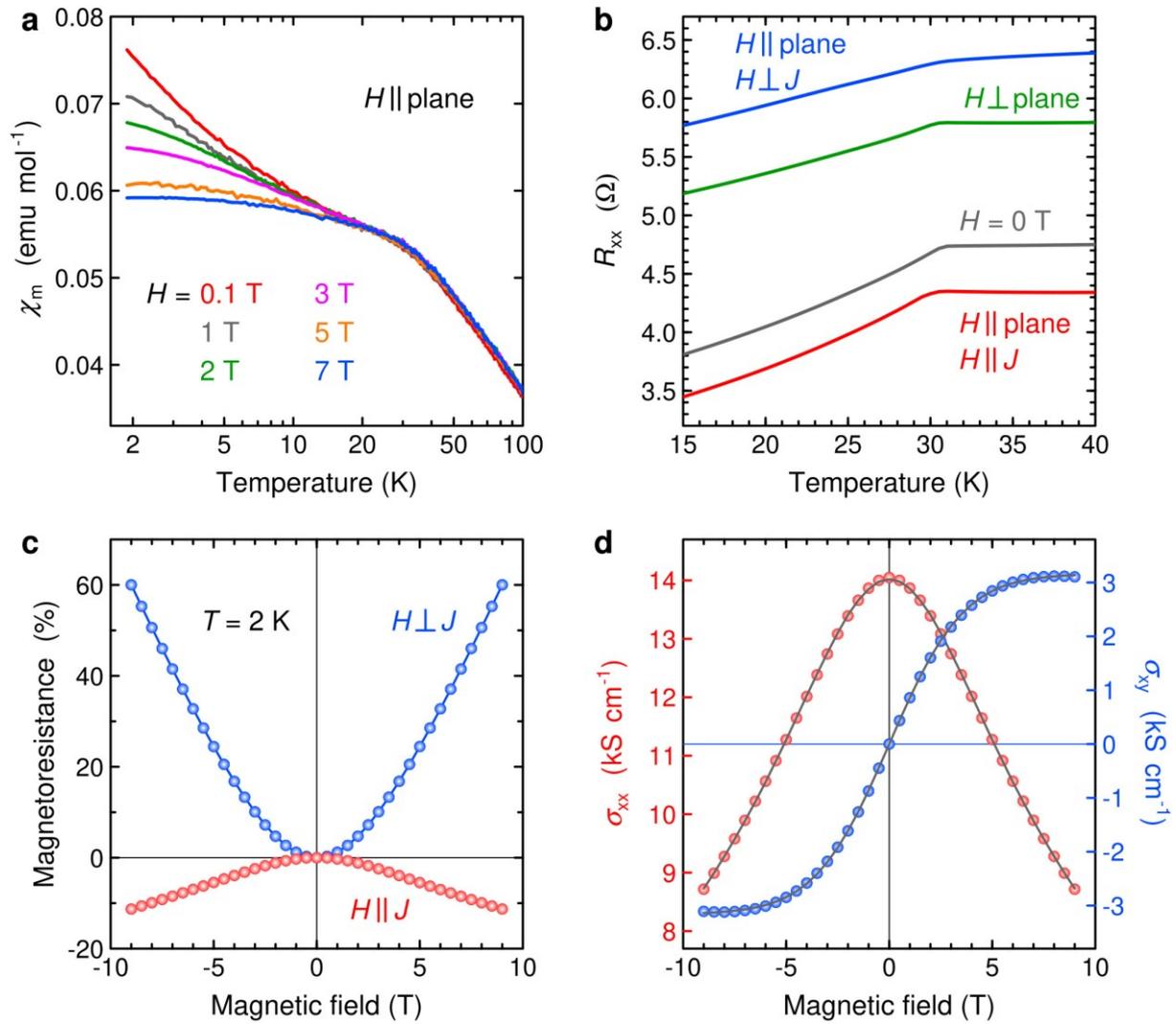

**Figure 2.** Magnetic and transport properties of a thick film ($d$ = 134 ML) of GdAlSi. a) Temperature dependence of the molar magnetic susceptibility $\chi_m$ in in-plane magnetic fields $H$ = 0.1 T (red), 1 T (grey), 2 T (green), 3 T (magenta), 5 T (orange), and 7 T (blue). b) Temperature dependence of resistance in zero magnetic field (grey), out-of-plane (green), in-plane parallel (red) and perpendicular (blue) to the current $J$ magnetic fields of $H$ = 9 T. c) MR at $T$ = 2 K in in-plane magnetic fields parallel (red) and perpendicular (blue) to the current $J$. d) Longitudinal (red) and transverse (blue) magnetoconductance at $T$ = 2 K; dots show experimental data while solids lines fit them by a three-band model.

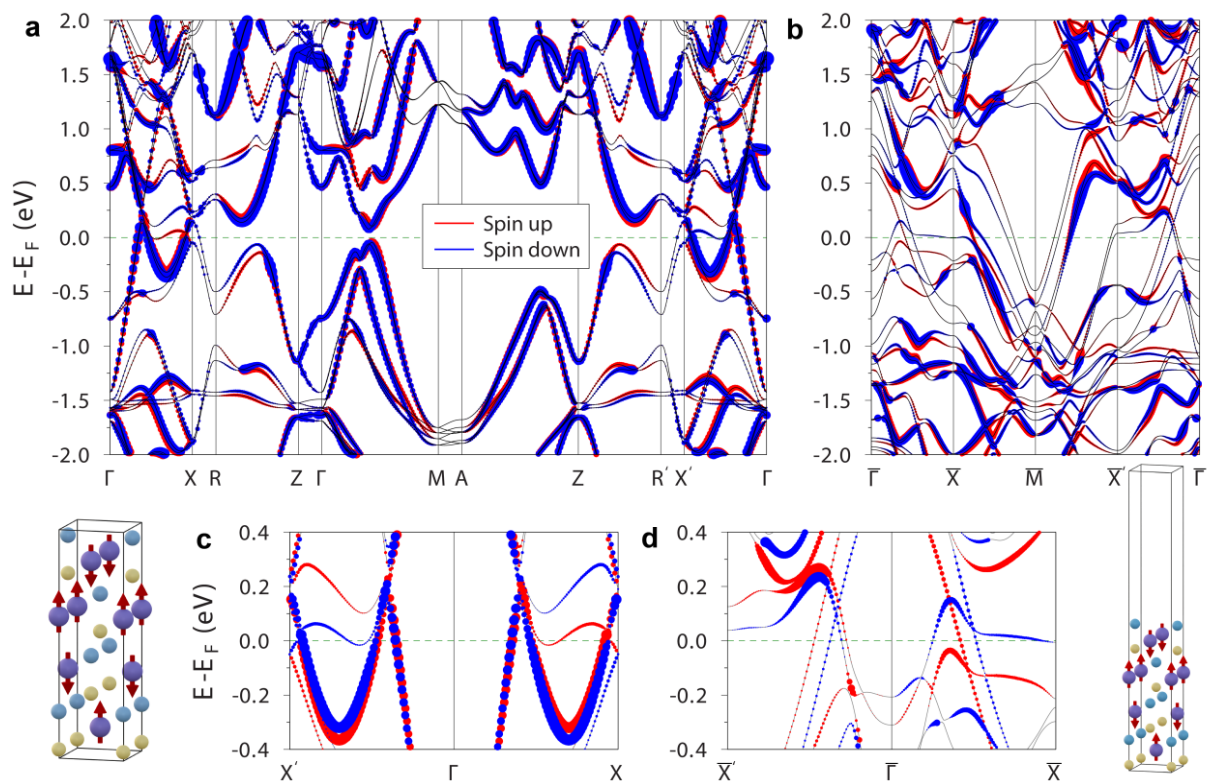

**Figure 3.** Calculated band structure of a) bulk and b) 1 ML GdAlSi. Expanded views of the band structures of c) bulk across the line $X' - \Gamma - X$ and d) 1 ML GdAlSi across the line $\overline{X}' - \overline{\Gamma} - \overline{X}$. The calculations employ the tetragonal conventional unit cell, in the case of 1 ML modified by adding a vacuum spacing. Bands with spins up and down are marked as red and blue, respectively.

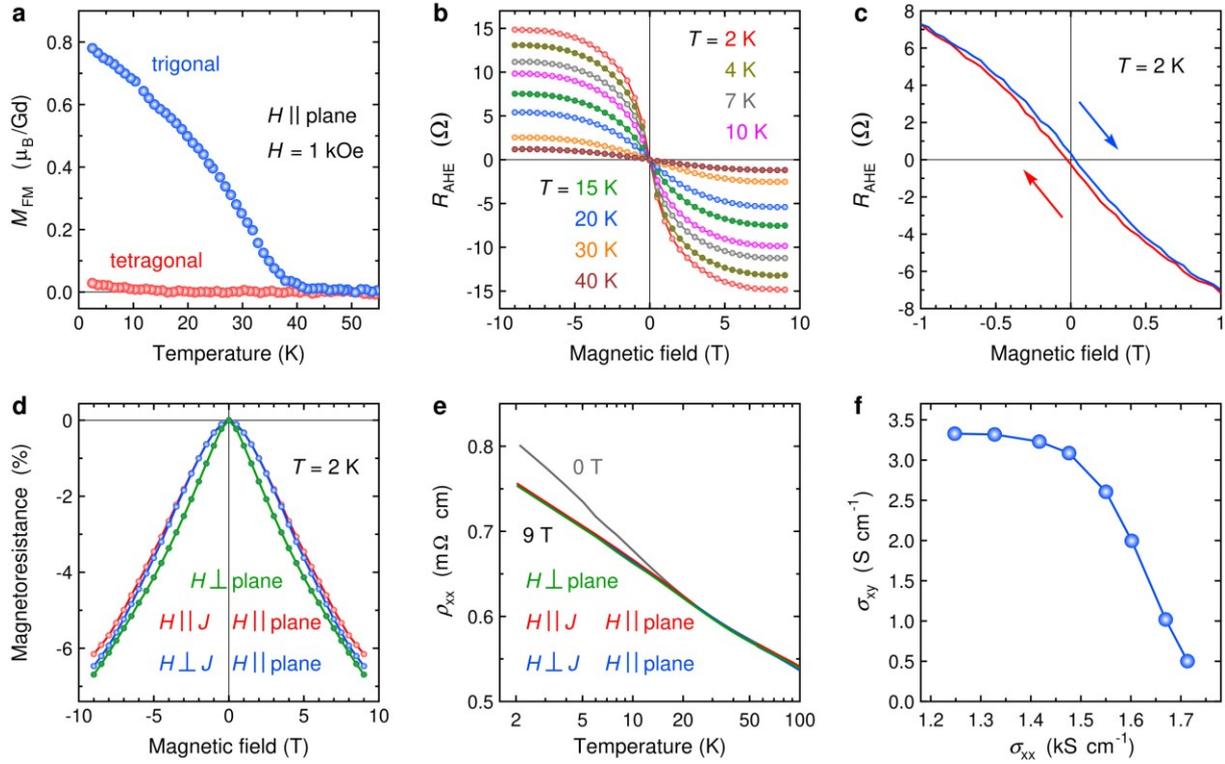

**Figure 4.** Magnetic and transport properties of 1 ML GdAlSi. a) Temperature dependence of the FM moment in in-plane magnetic field $H$ = 1 kOe: 1 ML of tetragonal GdAlSi (red) as compared to a film of trigonal GdAlSi (blue) of about the same thickness. b) AHE resistance at $T$ = 2 K (red), 4 K (olive), 7 K (grey), 10 K (magenta), 15 K (green), 20 K (blue), 30 K (orange), and 40 K (brown). c) Hysteresis in AHE resistance at $T$ = 2 K. d) MR at $T$ = 2 K for out-of-plane magnetic fields (green) and in-plane magnetic fields parallel (red) and perpendicular (blue) to the current $J$. e) Temperature dependence of resistivity in zero magnetic field (grey), out-of-plane (green), in-plane parallel (red) and perpendicular (blue) to the current $J$ magnetic field $H$ = 9 T. f) Dependence of the transverse (AHE) conductivity on the longitudinal conductivity.

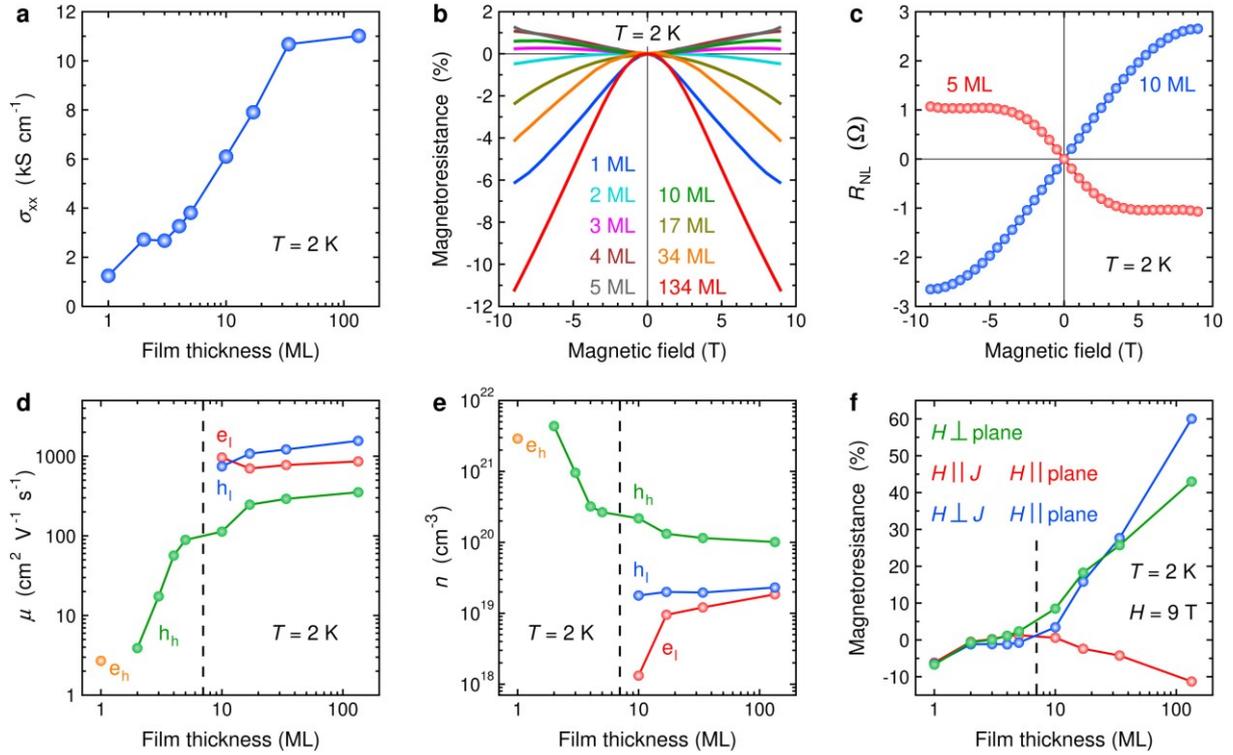

**Figure 5.** Thickness-dependent evolution of electron transport in GdAlSi. a) Dependence of longitudinal conductivity at $T = 2$ K on the film thickness. b) MR of GdAlSi at $T = 2$ K in in-plane magnetic fields parallel to the current $J$ for a selection of film thicknesses: 1 ML (blue), 2 ML (cyan), 3 ML (magenta), 4 ML (brown), 5 ML (grey), 10 ML (green), 17 ML (olive), 34 ML (orange), and 134 ML (red). c) Non-linear part of the Hall resistance at $T = 2$ K in 5 ML (red) and 10 ML (blue) GdAlSi. Film thickness dependence of d) carrier mobility and e) carrier concentration for bands of heavy holes ($h_h$, green), heavy electrons ($e_h$, orange), light holes ($h_l$, blue), and light electrons ($e_l$, red) at $T = 2$ K. f) Film thickness dependence of MR at $T = 2$ K in magnetic field $H = 9$ T: out-of-plane (green), in-plane parallel (red) and perpendicular (blue) to the current $J$. In d)–f), the dashed lines mark crossover to the 2D limit.

The Weyl altermagnet GdAlSi is brought to the ultimate 2D limit by epitaxy on Si; the evolution of its properties with the film thickness is studied. A fundamental transformation of the electron states accompanies the 3D-to-2D crossover. In particular, a spontaneous anomalous Hall effect emerges in ultra-thin films, attributed to crystal symmetry breaking.

O. E. Parfenov, D. V. Averyanov, I. S. Sokolov, A. N. Mihalyuk, O. A. Kondratev, A. N. Taldenkov, A. M. Tokmachev, V. G. Storchak*

**Pushing an Altermagnet to the Ultimate 2D Limit: Symmetry Breaking in Monolayers of GdAlSi**

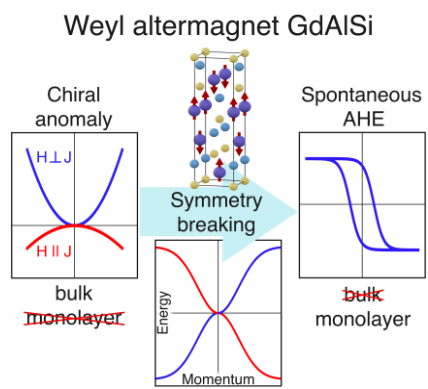

Supporting Information

**Pushing an Altermagnet to the Ultimate 2D Limit: Symmetry Breaking in Monolayers of GdAlSi**

*Oleg E. Parfenov, Dmitry V. Averyanov, Ivan S. Sokolov, Alexey N. Mihalyuk, Oleg A. Kondratev, Alexander N. Taldenkov, Andrey M. Tokmachev, and Vyacheslav G. Storchak\**

**Content:**

**Figure S1.** RHEED images of GdAlSi films of different thickness.

**Figure S2.** θ-2θ XRD scans of GdAlSi films of different thickness.

**Figure S3.** Thickness fringes in the XRD scan of 10 ML GdAlSi.

**Figure S4.** Temperature dependence of $\chi_m$ in 134 ML GdAlSi.

**Figure S5.** Curie-Weiss law in the paramagnetic region of 134 ML GdAlSi.

**Figure S6.** Shift by magnetic field of $R(T)$ at $T_N$ of 134 ML GdAlSi.

**Figure S7.** MR of 134 ML GdAlSi in the paramagnetic region.

**Figure S8.** Shubnikov-de Haas oscillations in 134 ML GdAlSi.

**Figure S9.** Carrier mobility and concentration in 134 ML GdAlSi.

**Figure S10.** Band structure of bulk GdAlSi.

**Figure S11.** Comparison of *M-H* loops in trigonal and tetragonal GdAlSi.

**Figure S12.** Temperature dependence of the AHE resistance in 1 ML GdAlSi.

**Figure S13.** MR of 1 ML GdAlSi at different temperatures.

**Figure S14.** Temperature dependence of resistivity in 3 and 34 ML GdAlSi.

**Figure S15.** MR of GdAlSi films in out-of-plane magnetic fields.

**Figure S16.** MR of GdAlSi films at different temperatures.

**Figure S17.** Temperature dependence of MR at 9 T in GdAlSi films.

**Figure S18.** AHE resistance of GdAlSi films at different temperatures.

**Figure S19.** MR anisotropy in GdAlSi films of different thickness.

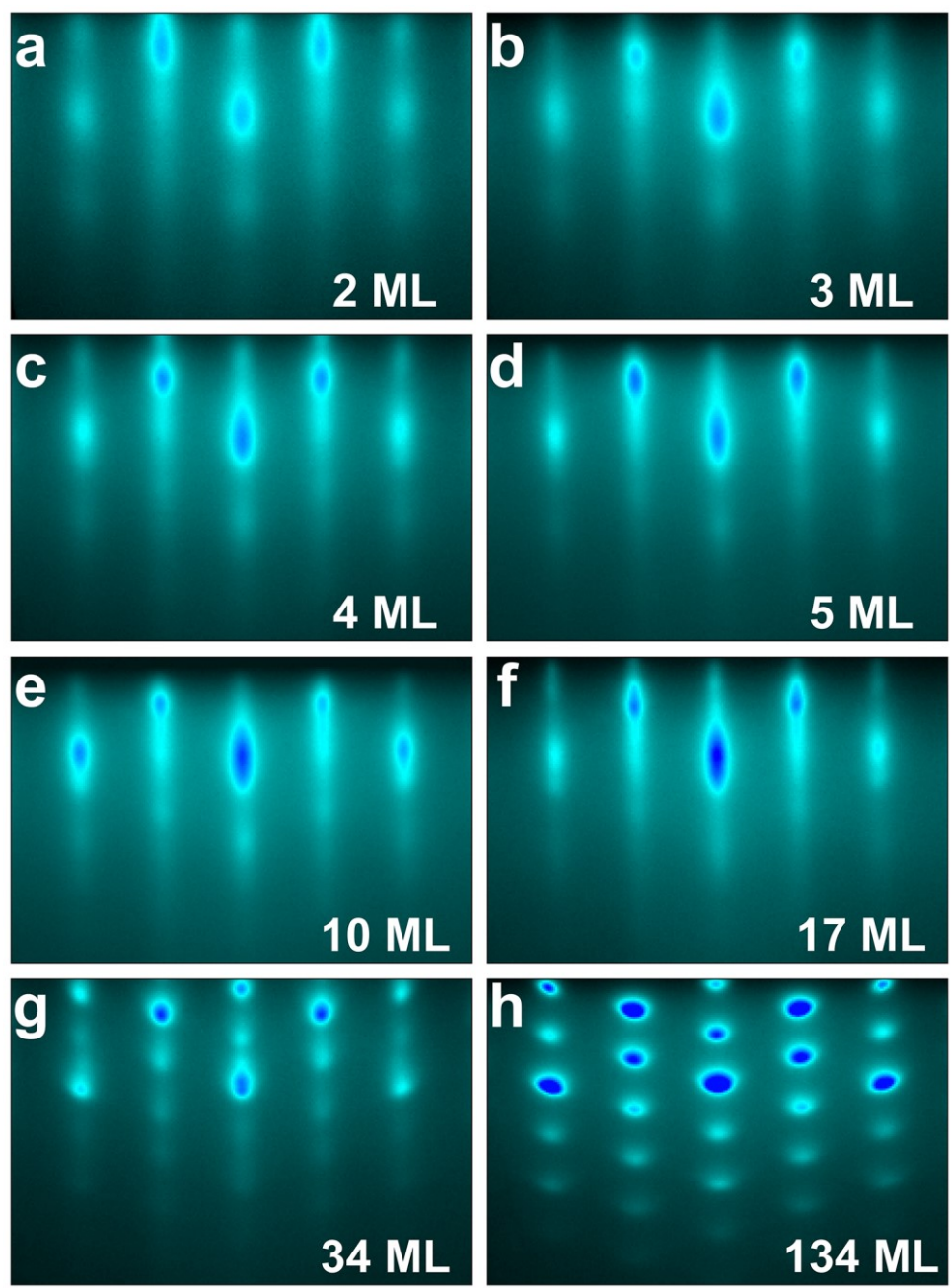

**Figure S1.** RHEED spectra of a) 2 ML, b) 3 ML, c) 4 ML, d) 5 ML, e) 10 ML, f) 17 ML, g) 34 ML, and h) 134 ML films of GdAlSi on Si(001) viewed along the [110] azimuth of the substrate.

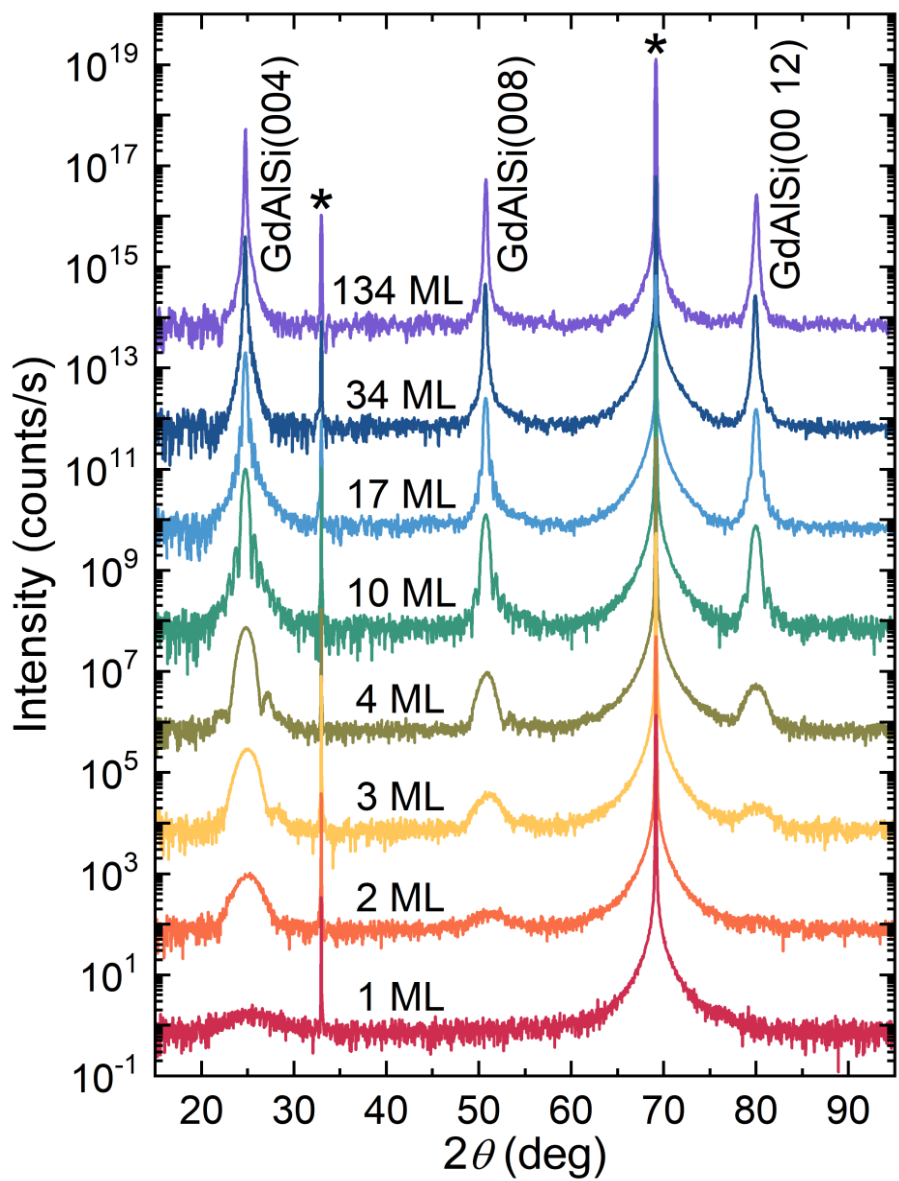

**Figure S2.** θ-2θ XRD scans of GdAlSi films on Si(001): 1 ML (red), 2 ML (orange), 3 ML (yellow), 4 ML (olive), 10 ML (green), 17 ML (light blue), 34 ML (dark blue), and 134 ML (purple). The curves are shifted to enhance the visibility of the peaks. The asterisks mark peaks from the Si(001) substrate.

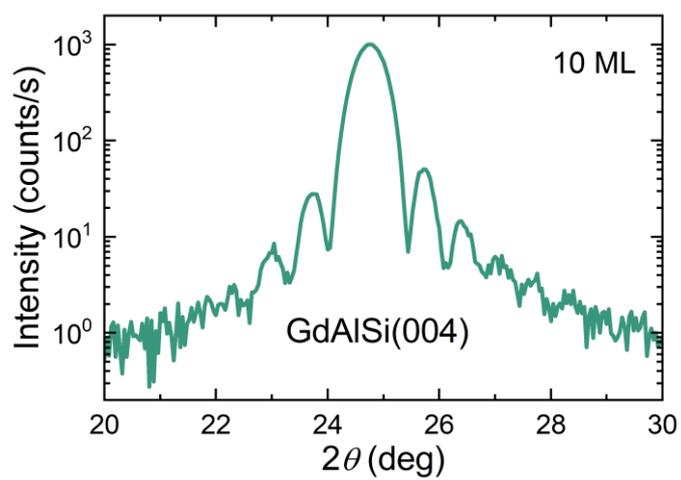

**Figure S3.** Thickness fringes around the (004) peak in the θ-2θ XRD scan of 10 ML GdAlSi.

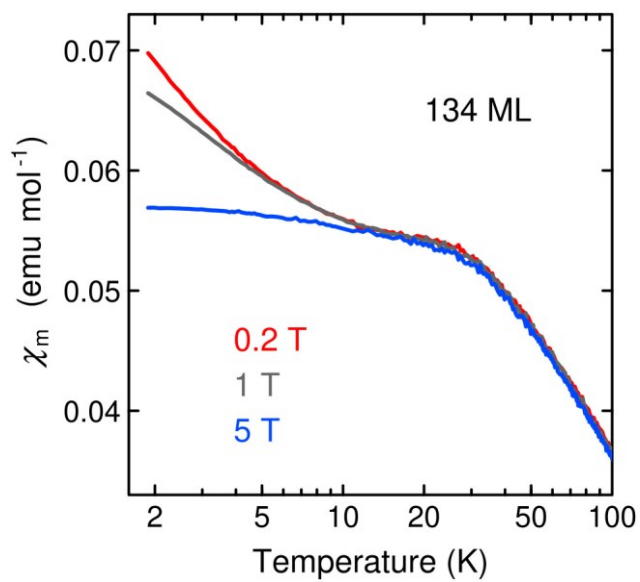

**Figure S4.** Temperature dependences of the molar magnetic susceptibility $\chi_m$ of 134 ML GdAlSi in out-of-plane magnetic fields $H$ = 0.2 T (red), 1 T (grey), and 5 T (blue).

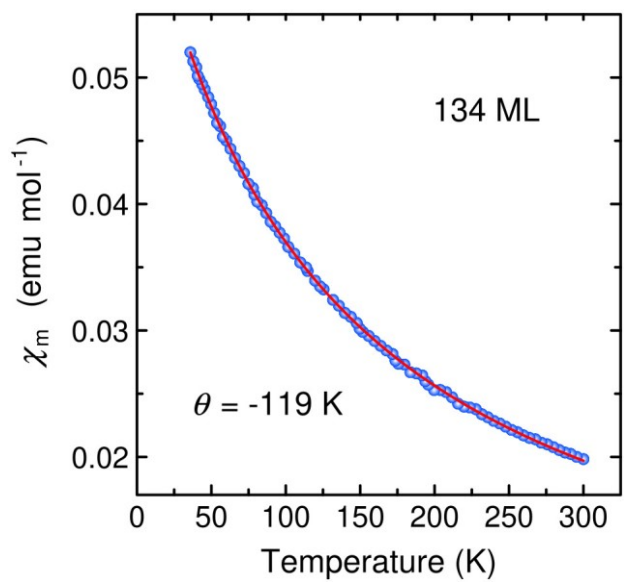

**Figure S5.** Temperature dependence of the molar magnetic susceptibility $\chi_m$ of 134 ML GdAlSi in an in-plane magnetic field $H = 1$ T (blue dots) and its fit by the Curie-Weiss law (red curve) with $\theta = -119$ K.

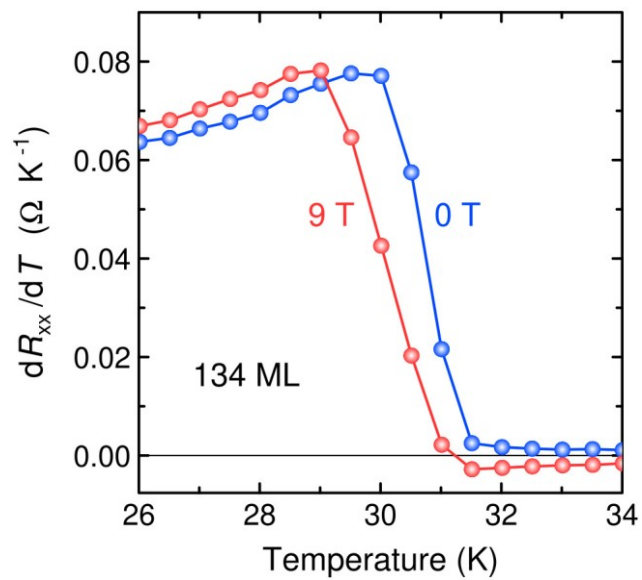

**Figure S6.** Temperature dependences of the temperature derivative of resistance around the magnetic transition in 134 ML GdAlSi, measured in zero magnetic field (blue) and in an in-plane magnetic field $H = 9$ T parallel to the current.

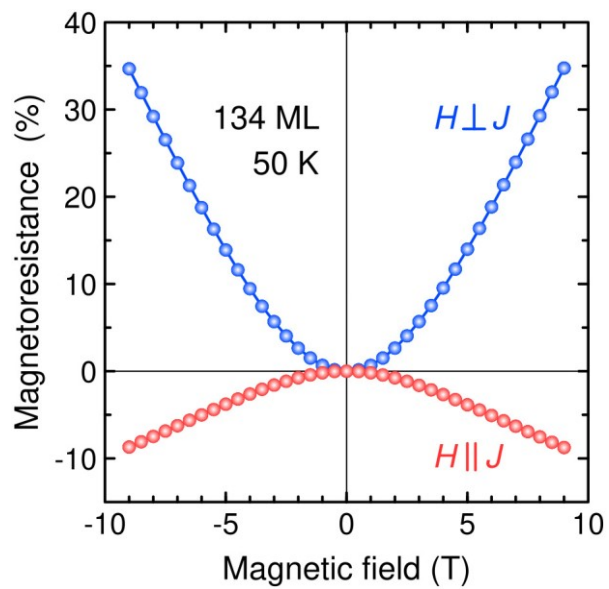

**Figure S7.** MR of 134 ML GdAlSi at $T = 50$ K in in-plane magnetic fields parallel (red) and perpendicular (blue) to the current $J$.

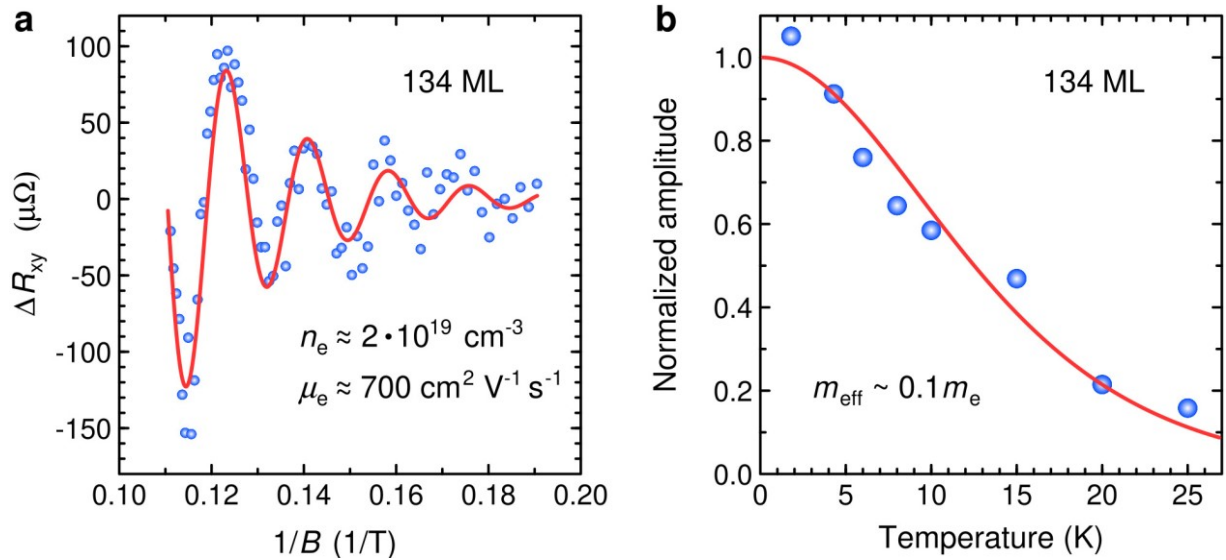

**Figure S8.** Shubnikov-de Haas oscillations in 134 ML GdAlSi. a) Non-linear part of the Hall resistance in out-of-plane magnetic fields at $T = 2$ K (blue dots) and its fit by the Lifshitz-Kosevich formalism (red line). b) Temperature dependence of the normalized amplitude of the oscillations (blue dots) and its fit to $cT/\sinh(cT)$ (red curve).

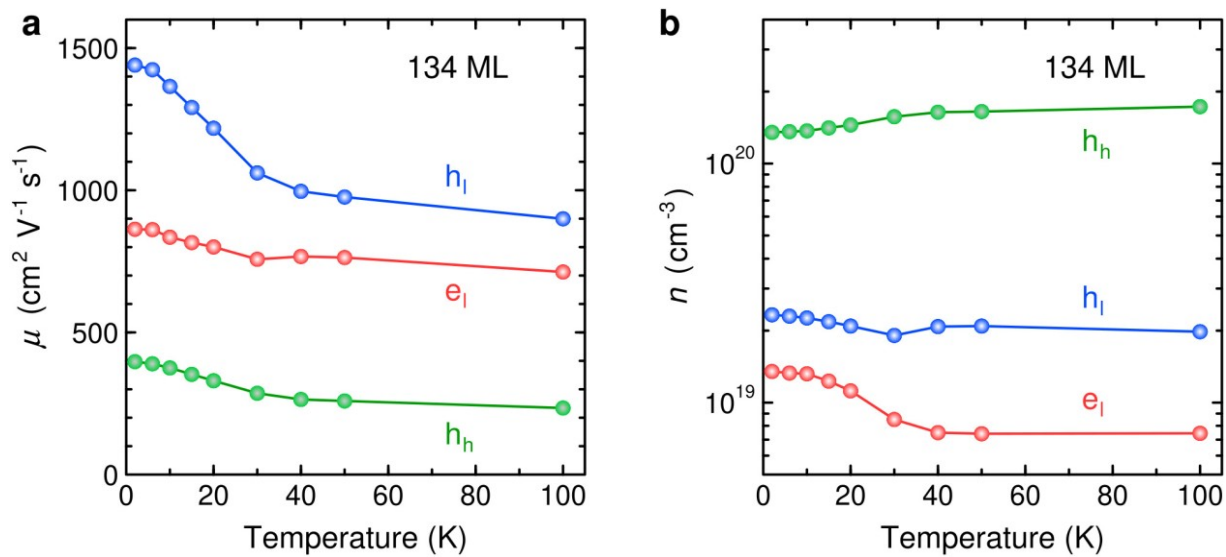

**Figure S9.** Temperature dependences of a) the carrier mobility and b) the carrier concentration for the bands of heavy holes ($h_h$, green), light holes ($h_l$, blue), and light electrons ($e_l$, red) in the three-band model of 134 ML GdAlSi.

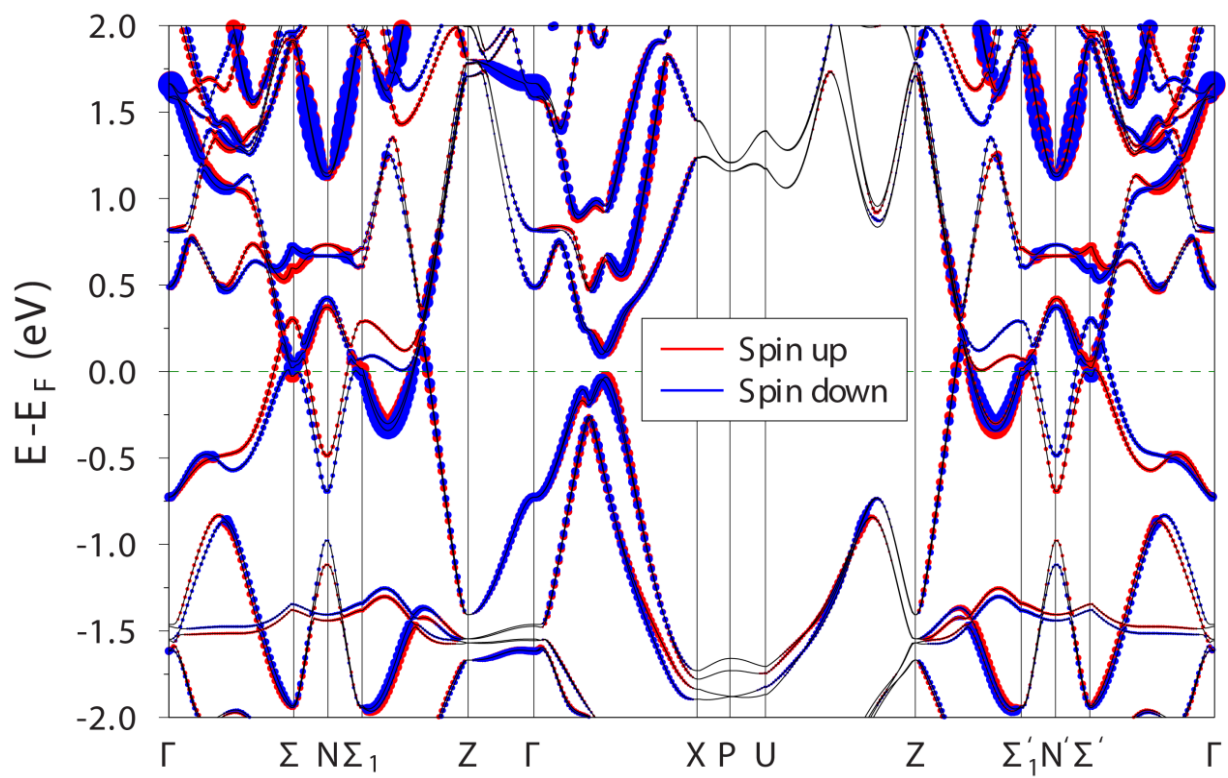

**Figure S10.** Calculated band structure of bulk GdAlSi demonstrating NRSS. Bands with spins up and down are marked as red and blue, respectively. The image and notation correspond to the primitive body-centered tetragonal unit cell (*cf.* Figure S5 of Ref. [48]).

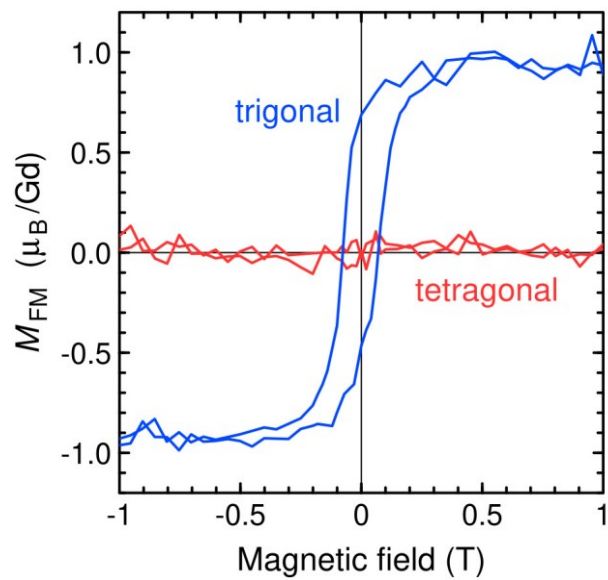

**Figure S11.** Magnetic field dependences of the magnetic moment at $T$ = 2 K: 1 ML of tetragonal GdAlSi (red) as compared to a film of trigonal GdAlSi (blue) of about the same thickness.

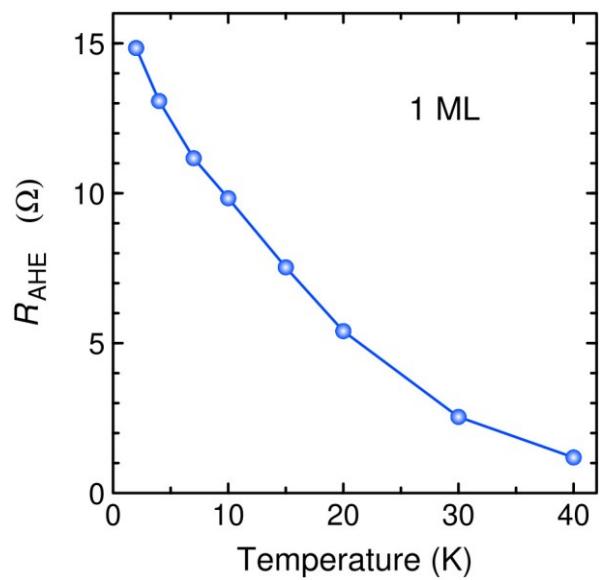

**Figure S12.** Temperature dependence of the AHE resistance at $H$ = -9 T in 1 ML GdAlSi.

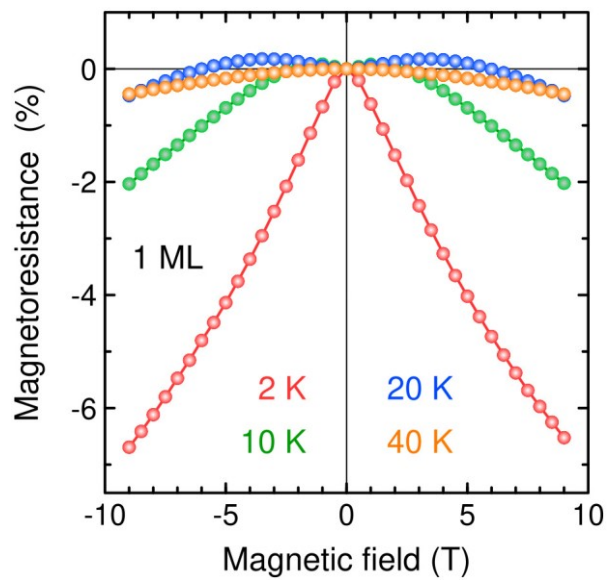

**Figure S13.** MR of 1 ML GdAlSi at $T$ = 2 K (red), 10 K (green), 20 K (blue), and 40 K (orange).

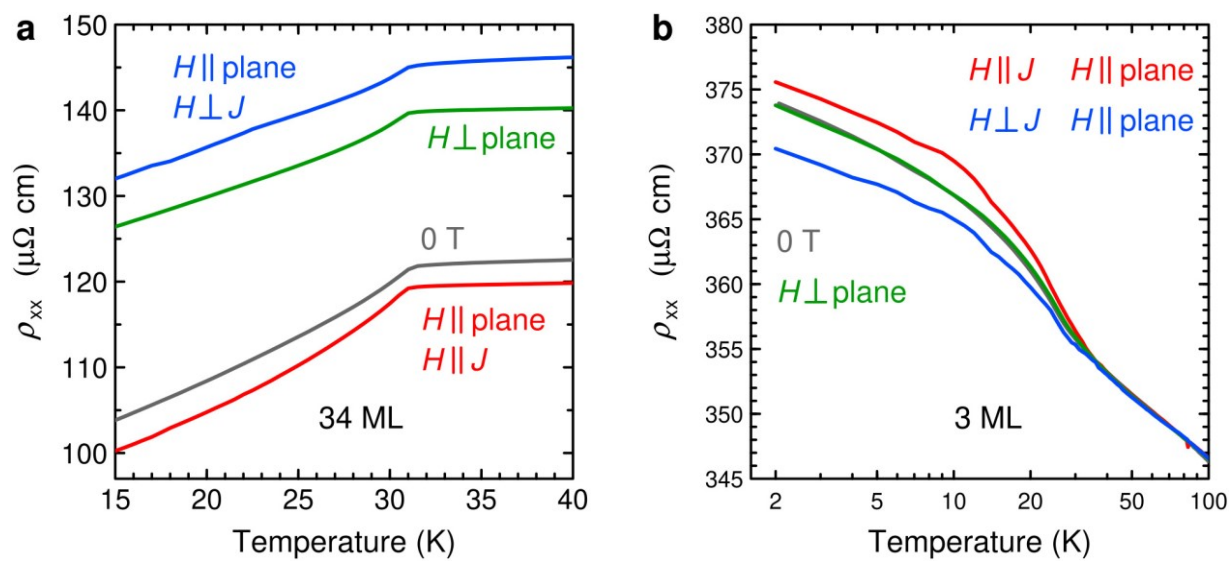

**Figure S14.** Temperature dependences of resistivity in zero magnetic field (grey), out-of-plane (green), in-plane parallel (red) and perpendicular (blue) to the current *J* magnetic field $H = 9$ T: a) 34 ML GdAlSi and b) 3 ML GdAlSi.

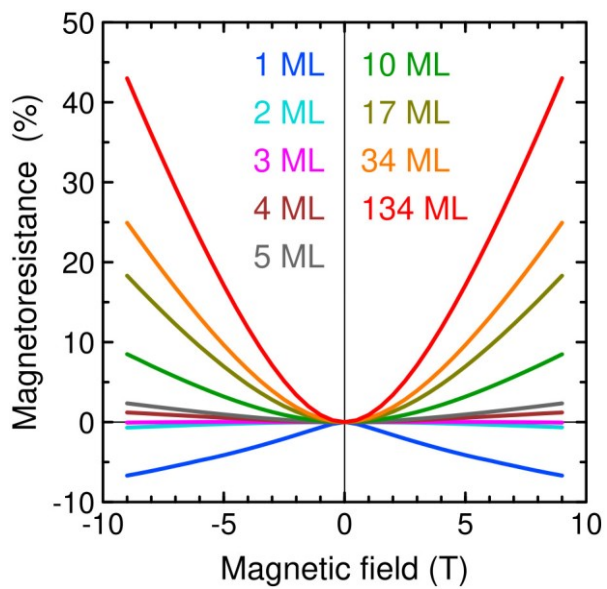

**Figure S15.** MR of GdAlSi at $T = 2$ K in out-of-plane magnetic fields for a selection of film thicknesses: 1 ML (blue), 2 ML (cyan), 3 ML (magenta), 4 ML (brown), 5 ML (grey), 10 ML (green), 17 ML (olive), 34 ML (orange), and 134 ML (red).

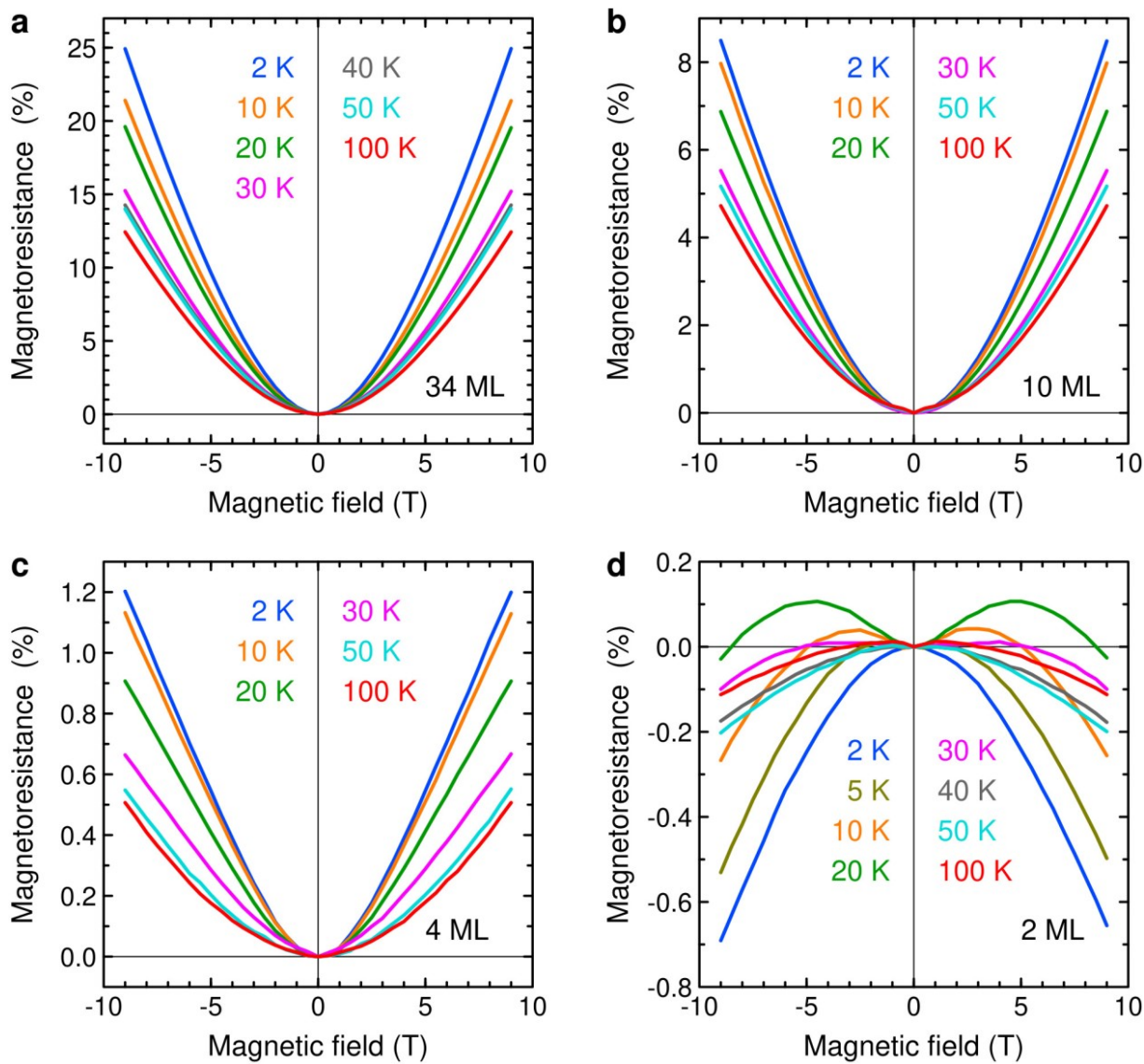

**Figure S16.** MR of GdAlSi in out-of-plane magnetic fields for a selection of film thicknesses: a) 34 ML at $T$ = 2 K (blue), 10 K (orange), 20 K (green), 30 K (magenta), 40 K (grey), 50 K (cyan), and 100 K (red); b) 10 ML at $T$ = 2 K (blue), 10 K (orange), 20 K (green), 30 K (magenta), 50 K (cyan), and 100 K (red); c) 4 ML at $T$ = 2 K (blue), 10 K (orange), 20 K (green), 30 K (magenta), 50 K (cyan), and 100 K (red); d) 2 ML at $T$ = 2 K (blue), 5 K (olive), 10 K (orange), 20 K (green), 30 K (magenta), 40 K (grey), 50 K (cyan), and 100 K (red).

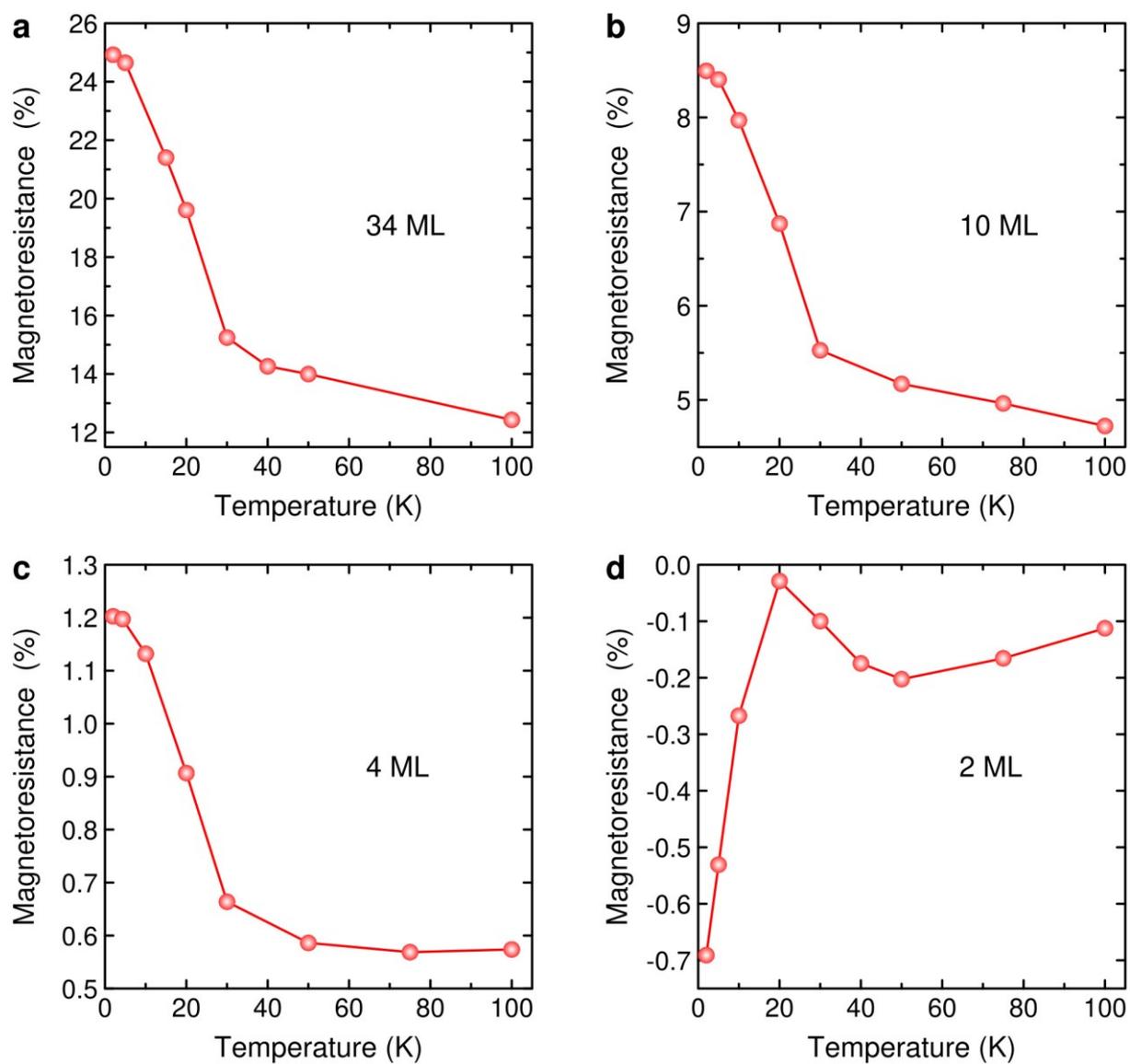

**Figure S17.** Temperature dependences of MR in out-of-plane magnetic field $H$ = 9 T for GdAlSi films of a) 34 ML, b) 10 ML, c) 4 ML, and d) 2 ML thickness.

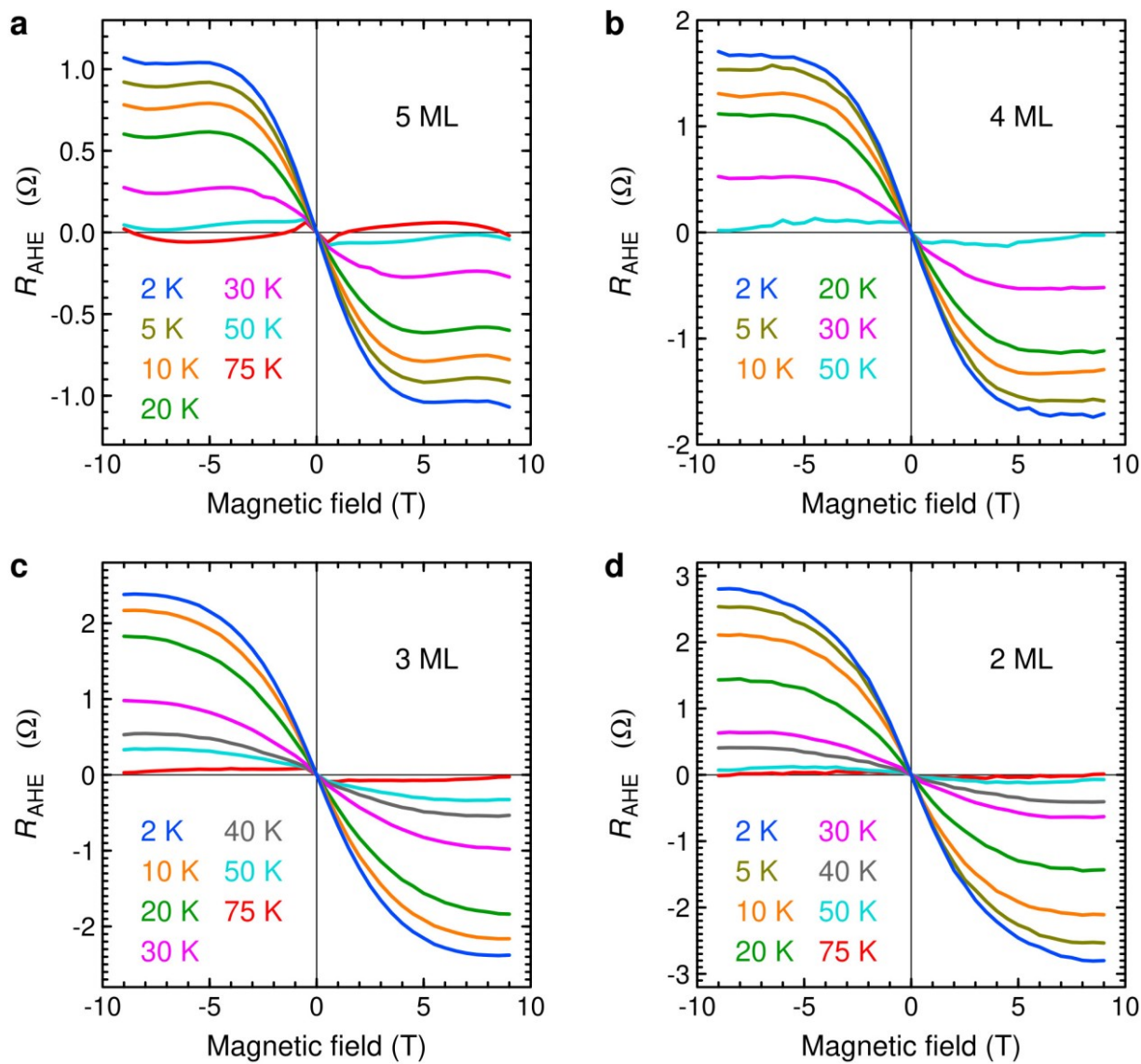

**Figure S18.** AHE resistance in GdAlSi for a selection of film thicknesses: a) 5 ML at $T = 2$ K (blue), 5 K (olive), 10 K (orange), 20 K (green), 30 K (magenta), 50 K (cyan), and 75 K (red); b) 4 ML at $T = 2$ K (blue), 5 K (olive), 10 K (orange), 20 K (green), 30 K (magenta), and 50 K (cyan); c) 3 ML at $T = 2$ K (blue), 10 K (orange), 20 K (green), 30 K (magenta), 40 K (grey), 50 K (cyan), and 75 K (red); d) 2 ML at $T = 2$ K (blue), 5 K (olive), 10 K (orange), 20 K (green), 30 K (magenta), 40 K (grey), 50 K (cyan), and 75 K (red).

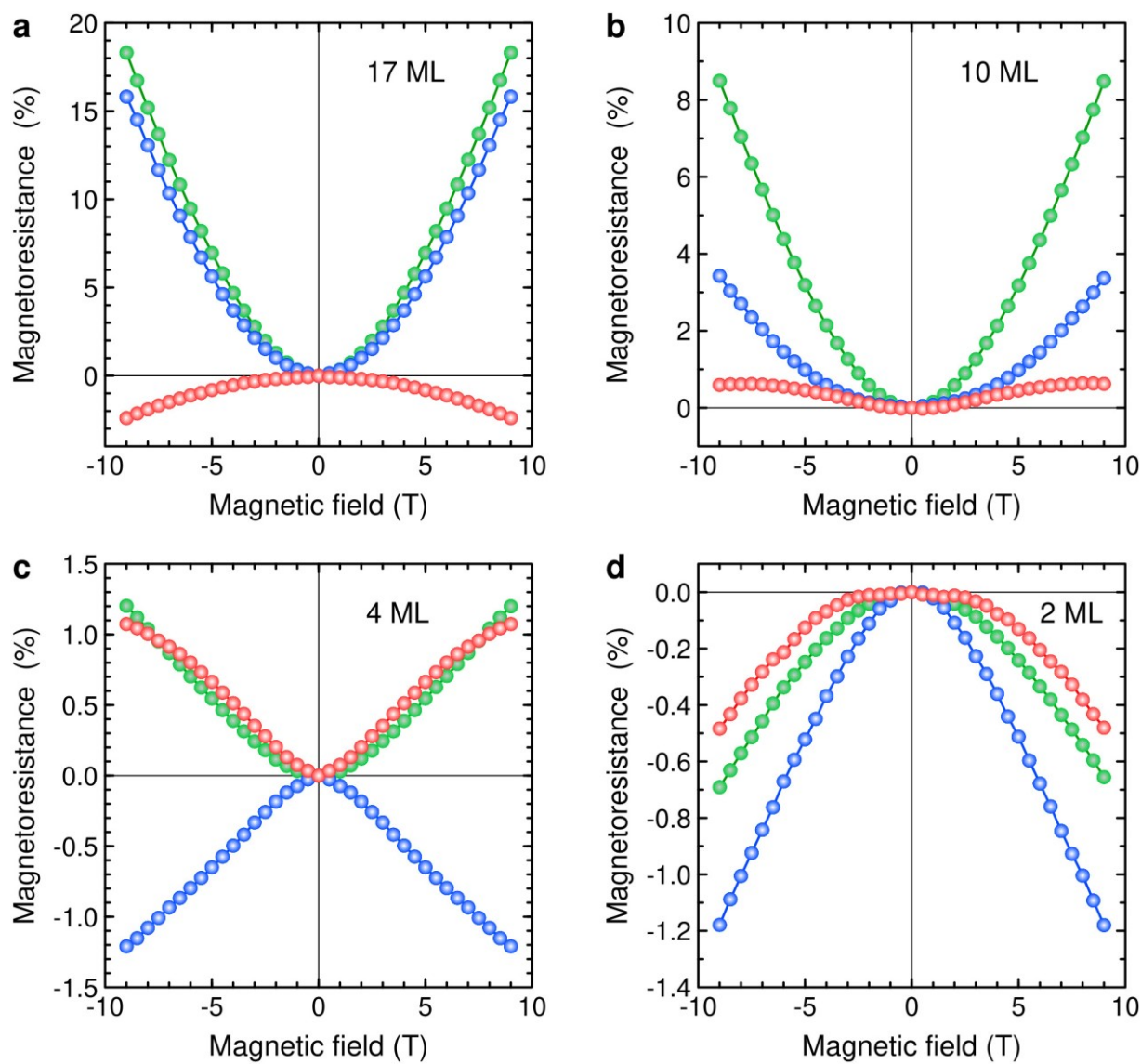

**Figure S19.** MR at $T = 2$ K in a) 17 ML, b) 10 ML, c) 4 ML, and d) 2 ML GdAlSi films in out-of-plane magnetic fields (green) and in-plane magnetic fields parallel (red) and perpendicular (blue) to the current $J$.